\documentclass[preprint,12pt,3p]{elsarticle}

\usepackage{graphicx,color}
\usepackage{epsfig}

\usepackage{amssymb}
\usepackage{hyperref}

\journal{Manufacturing Letters}

\begin{document}

\begin{frontmatter}

\title{Efficient planning of peen-forming patterns via artificial neural networks}

\author[label1]{Wassime Siguerdidjane}
\address[label1]{Department of Mechanical Engineering, Polytechnique Montr\'eal, Canada}
\address[label2]{Aluminium Research Centre – REGAL, Canada}

\author[label1,label2]{Farbod Khameneifar\corref{cor1}}
\ead{farbod.khameneifar@polymtl.ca}
\cortext[cor1]{Corresponding author}

\author[label1,label2]{Fr\'ed\'erick P. Gosselin}

\begin{abstract}
Robust automation of the shot peen forming process demands a closed-loop feedback in which a suitable treatment pattern needs to be found in real-time for each treatment iteration. In this work, we present a method for finding the peen-forming patterns, based on a neural network (NN), which learns the nonlinear function that relates a given target shape (input) to its optimal peening pattern (output), from data generated by finite element simulations. The trained NN yields patterns with an average binary accuracy of 98.8\% with respect to the ground truth in microseconds.
\end{abstract}

\end{frontmatter}


\section{Introduction}
\label{sec1}
The performance of artificial neural networks has been proven on visual pattern recognition tasks \cite{krizhevsky_imagenet_2012}. Here, we demonstrate that the inverse problem of shot peen forming, consisting in finding a suitable treatment, can be formulated as a visual pattern recognition problem, and accurately and efficiently solved using a neural network.

Peen forming is the process of shaping thin metal parts by bombarding them with a stream of shot at high velocity. Treating certain regions of a part gives rise to local curvature, without expensive geometry-specific tools, such as dies and matrices.  This process is frequently used in the aerospace industry to correct distortion in machined parts and to form large panels such as wing skins \cite{kulkarni_investigation_1981}.

The final deformed shape of the part depends on the treated regions, referred to as the peening pattern. To select the peening pattern, today's industry mostly relies on human intuition with operators trained in this craft with a qualitative understanding of the mechanics through trial and error. This is time-consuming, inaccurate and expensive. 
 
\begin{figure}[ht]
     \centering
         \includegraphics[width=\textwidth]{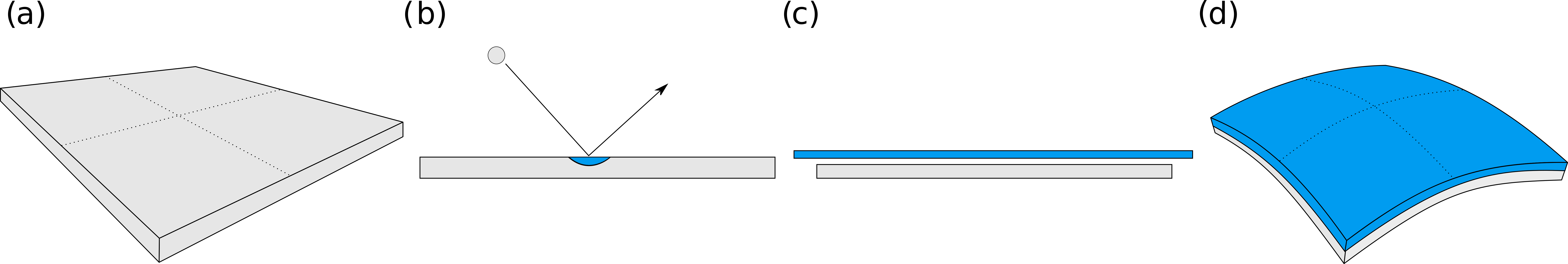}
         \caption{Representing the mechanics of peen forming as a bilayer plate problem: (a) Initially flat plate; (b)  Visualization of the local effect of an impact; (c) Multiple impacts expand the upper portion of the plate, while the rest of the plate is unaffected, causing a geometrical incompatibility; (d) Final deformed plate uniformly treated. Inspired by Ref. \cite{thesisPierre}.}
    \label{fig:cartoon}
\end{figure}

 Predictive models are crucial to predict the outcome of a planned peening pattern. Different modelling approaches have been proposed for peen forming \cite{xiao_prediction_2019,vanluchene_numerical_1996,gariepy_shot_2011}. Notably, Faucheux et al. \cite{FAUCHEUX2018135} proposed the bilayer framework. 
Fig. \ref{fig:cartoon} depicts how the cumulative effect of multiple shot impacts induce surface compressive residual stresses, which in turn result in the local expansion of the treated shallow portion of the plate. Combining the treated expanded layer with the unaffected bottom layer forms a curved bilayer. Thermal expansion efficiently models the effect of the treatment, and is readily available in most finite element software. Adjusting the thermal expansion coefficient and the thickness of the active layer controls the intensity of the process. 

The shape of real peen-formed parts still deviates from the model's predicted shape, mainly due to unknown initial residual stresses in parts \cite{thesisPierre}. Since the measurement of residual stresses is either expensive or destructive \cite{stress}, a potential solution is to adjust the model using feedback during the process. The long-term objective of our research is to maximize the robustness of automated peen-forming by incorporating modelling and 3D shape measurement in the control loop for adaptive adjustment of the model with the feedback from shape measurement during peen forming.

The bilayer framework efficiently simulates the deformed shape when applying a peening pattern, i.e., the direct problem. However, the inverse problem can only be solved through iterative optimization.
 The difficulty is that the direct problem should be solved in every iteration of the optimization, which makes it inefficient to solve the inverse problem. For closed-loop process control, the inverse problem needs to be solved for each treatment iteration. The development of a more efficient method for real-time process planning is therefore necessary.


Here we demonstrate that with the bilayer framework, predicting the peening treatment to form a flat plate into a target shape can be formulated as a pattern recognition problem, solvable with a trained neural network with sufficient accuracy, while only requiring a computation time of the order of microseconds.

\section{Theoretical Background on Neural Networks}
\label{sec2}
Neural Networks (NNs) are nonlinear function approximators optimized through gradient-based methods to fit a given dataset. NNs learn to predict the most probable output given the training data provided. They learn from known references (ground truths). The conventional training framework requires the dataset to be separated into a training set to optimize the network’s parameters, a validation set to validate and tune the network’s hyper-parameters, such as the network’s depth and learning rate, and a test set isolated to evaluate the network’s performance on unseen data. This training framework ensures that the model is not overfitting the training data \cite{Goodfellow-et-al-2016}.

\section{Methodology}
\label{sec3}
We propose to train a NN on data generated with the Finite Element Method (FEM) based on the bilayer framework \cite{FAUCHEUX2018135}. The objective of this NN is to predict the required peening pattern, when given a map of the average Gaussian curvature of each element. 

Fig. 2 shows the data generation pipeline. 60,000 examples were generated. 40,000 examples were used for training the NN, while 10,000 were used for validation. The remaining 10,000 examples from the dataset served to test the final performance of the model.

Since symmetry conditions are applied during the simulation, only a quarter of the plate is considered. The program generates the training patterns by initially creating random maze images of the size of a quarter of the plate (Fig. 2a) using the depth-first search algorithm \cite{algos} (adapted from \cite{maze_gen}).  The $A^*$ search algorithm \cite{algos} (adapted from \cite{maze_solver}) then finds a path within the maze to navigate the workpiece from one randomly selected point to another, hence creating a random continuous path (Fig. 2b). The path is then widened by recursively padding it with one pixel on each side. This operation ensures that the generated pattern is larger than a single pixel, which corresponds to the resolution of the mesh. This work assumes that the peen forming patterns considered would be experimentally enforced through the use of shot peening masks, and therefore are not limited by the diameter of the nozzle. All pixels on the generated path are assigned a value of 1 representing the requirement of treatment (blue pixels in Fig. 2c). Fig. 2d presents the final pattern image after being symmetrized to cover the full plate. The generation of these random patterns offers a diversity of continuous shapes that span the surface of the plate, to represent arbitrary reachable peen forming cases. 

The open-source FEM software CalculiX 
\cite{dhondt_finite_2004} then computes the deformed shape (Fig. 2f) of the initially flat plate with a static analysis considering geometric nonlinearities. The pattern is applied on the plate at a single time to represent the use of a mask. The bilayer model considers a real plate treated in the same way to be held flat during the peen forming process, and released to observe the deformation induced by the treatment.   

The square plate of size $L\times L\times h$ (32 mm $\times$ 32 mm $ \times$ 1 mm) is discretized on a grid of  32x32 S8R elements. These elements define a composite of two layers: the active top layer of thickness $h_{active}=0.1$ mm with a positive thermal expansion coefficient $\alpha$, and the passive bottom layer of thickness $h_{passive}=0.9$ mm, which is unattainable by the treatment.
A constant temperature $T=1$ $^{\circ}$C is applied. Increasing the $\alpha$ value of the treated elements applies the load on the plate. This simulates a bi-axial expansion $\epsilon=\alpha T$ in the active layer of the treated elements. This study considered $1.36 \mathrm{e}{-2} \leq \alpha \leq 1.09 \mathrm{e}{-1}$ $^{\circ}$C$^{-1}$. As in Ref. \cite{FAUCHEUX2018135}, the dimensionless bending moment generated by the expansion of the active layer is defined as 
\begin{equation}\label{eq:1}
    \mathit{\Gamma} = 6\left(1+\nu\right)\epsilon\frac{ h_{active} h_{passive} L^2}{h^4}
\end{equation}
where $\nu=0.33$ is the Poisson coefficient of the material of the plate. The process-oriented interpretation of $\mathit{\Gamma}$ defines the dimensionless peening intensity.
 
 Since the peen forming effect is predominantly bending, curvature is selected as an efficient representation of the plate deformation rather than nodal deflection. The Gaussian curvature is computed at each node of the mesh with the Gauss-Bonnet theorem implemented in the python library PyMesh \cite{pymesh}, and then averaged for each element. This shapes the data as two images of the same size, encoding respectively the target shape (Fig. 2f) and its corresponding pattern (Fig. 2d). The neural network takes as input the target curvature map and outputs the most probable peening pattern to reach it. The output pixel values correspond to the probability that an element should be treated, based on the statistical distribution of training data.

\begin{figure}[ht]
     \centering
         \centering
         \includegraphics[width=0.65\textwidth]{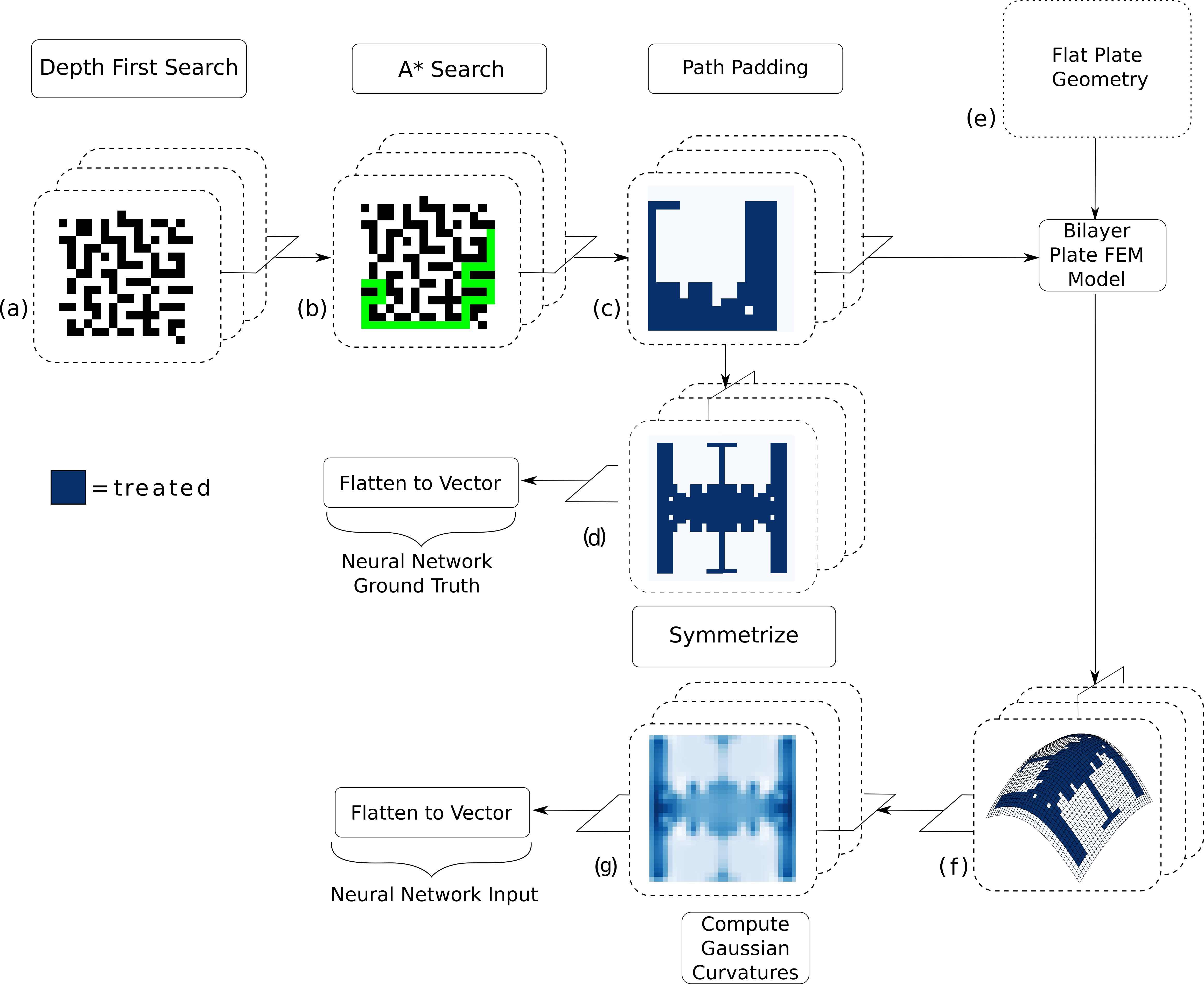}
         \caption{ Data generation pipeline: (a) Random maze image where each pixel represents an element of the FEM; (b) Continuous path in the maze found by the $A^*$ search algorithm; (c) Resulting peening pattern after path padding operations used to widen the peening area. This pattern covers a quarter of the plate, since symmetry conditions are leveraged in the FEM simulation; (d) The final pattern image used as ground truth, after being symmetrized to cover the full plate;  (e) Initially flat mesh of a quarter of the plate; (f) Deformed mesh resulting from the FEM bilayer simulation using the generated peening pattern; (g) Gaussian curvature image.
}
     \centering
\end{figure}

\begin{figure}
     \centering
         \centering
         \includegraphics[scale=0.6]{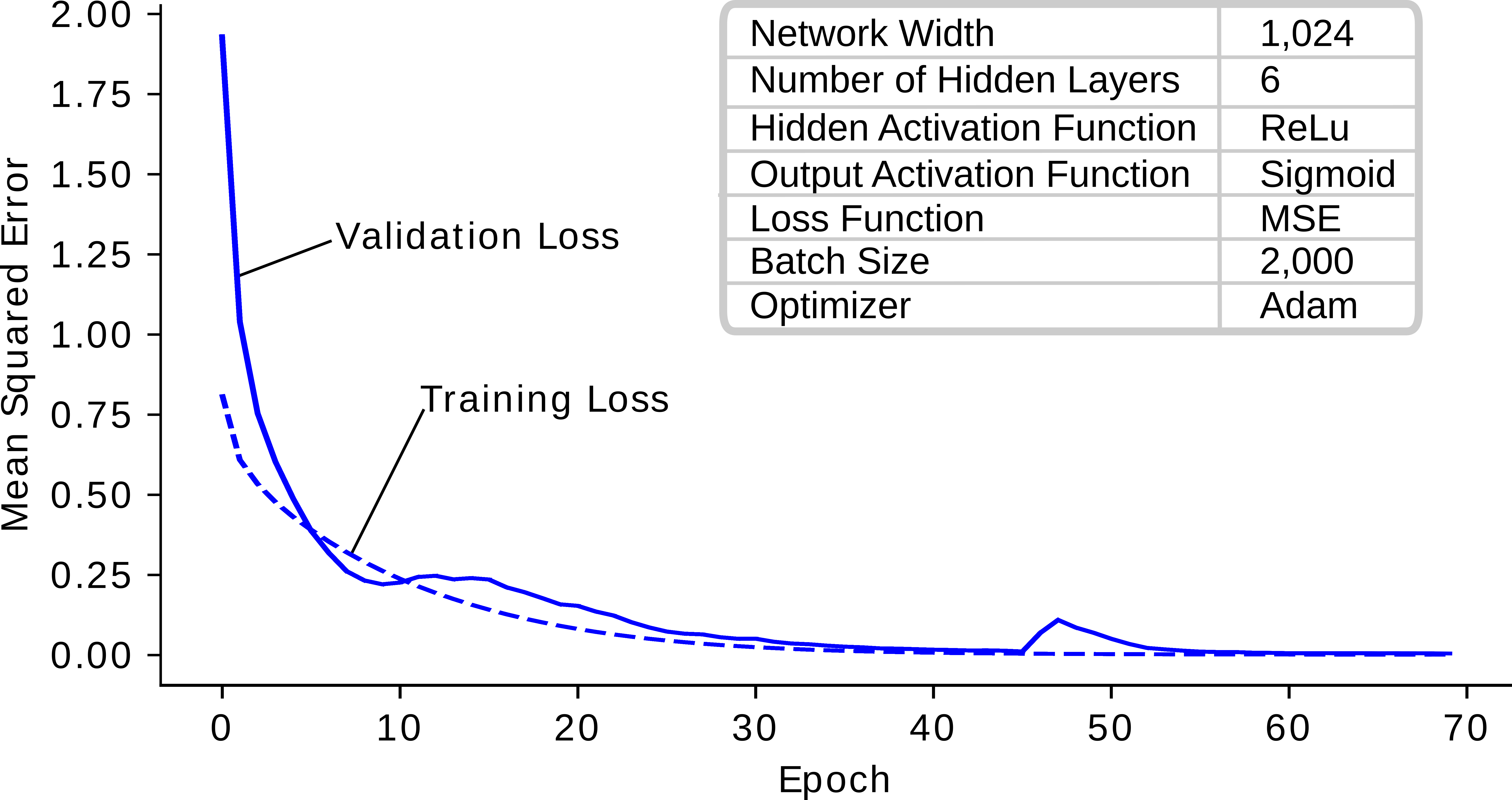}
         \caption{\label{fig:learning}Learning curve and hyper-parameters.}
     \centering
\end{figure}

Although a convolutional neural network \cite{Lenet} would have been an alternative, this work demonstrates that the simpler multilayer perceptron (MLP) is adequate for the current task. The number of elements in the plate mesh defines the width of the proposed six-layer MLP \cite{10.1007/978-3-642-70911-1_20}. The images in the dataset are flattened into vectors to be passed to the MLP. The model is implemented with Keras 2.2.4 \cite{chollet2015keras} and Tensorflow 1.12.0 \cite{tensorflow2015-whitepaper} as backend. Batch Normalization \cite{DBLP:journals/corr/IoffeS15} is applied to every layer. A grid search led to the model hyper-parameters selection (Fig. 3). The Adam optimizer \cite{kingma_adam_2017} executed the training. Backpropagation algorithm \cite{rumelhart_learning_1986} minimized the mean squared error (MSE) loss function between the predicted patterns and the known reference patterns, until the training loss and validation loss converged (Fig. \ref{fig:learning}). The MSE loss function was chosen empirically by comparing its performance with other loss functions such as cross-entropy.

\section{Results and Discussion}
\label{sec4}
The data generation took 20 hours, and the model training took 15 minutes. The prediction time of the model only takes 45 microseconds. This last value is crucial when considering the application of this model to a real-time process. The development and the evaluation of the model were conducted on a laptop computer with an Intel® Core™ i7-6700HQ CPU, with 16GB of RAM and under the Ubuntu 18.04 operating system. Graphics Processing Unit (GPU) acceleration was not leveraged in this study.

\begin{figure}[!ht]
    \centering
    \includegraphics[width=0.6\textwidth]{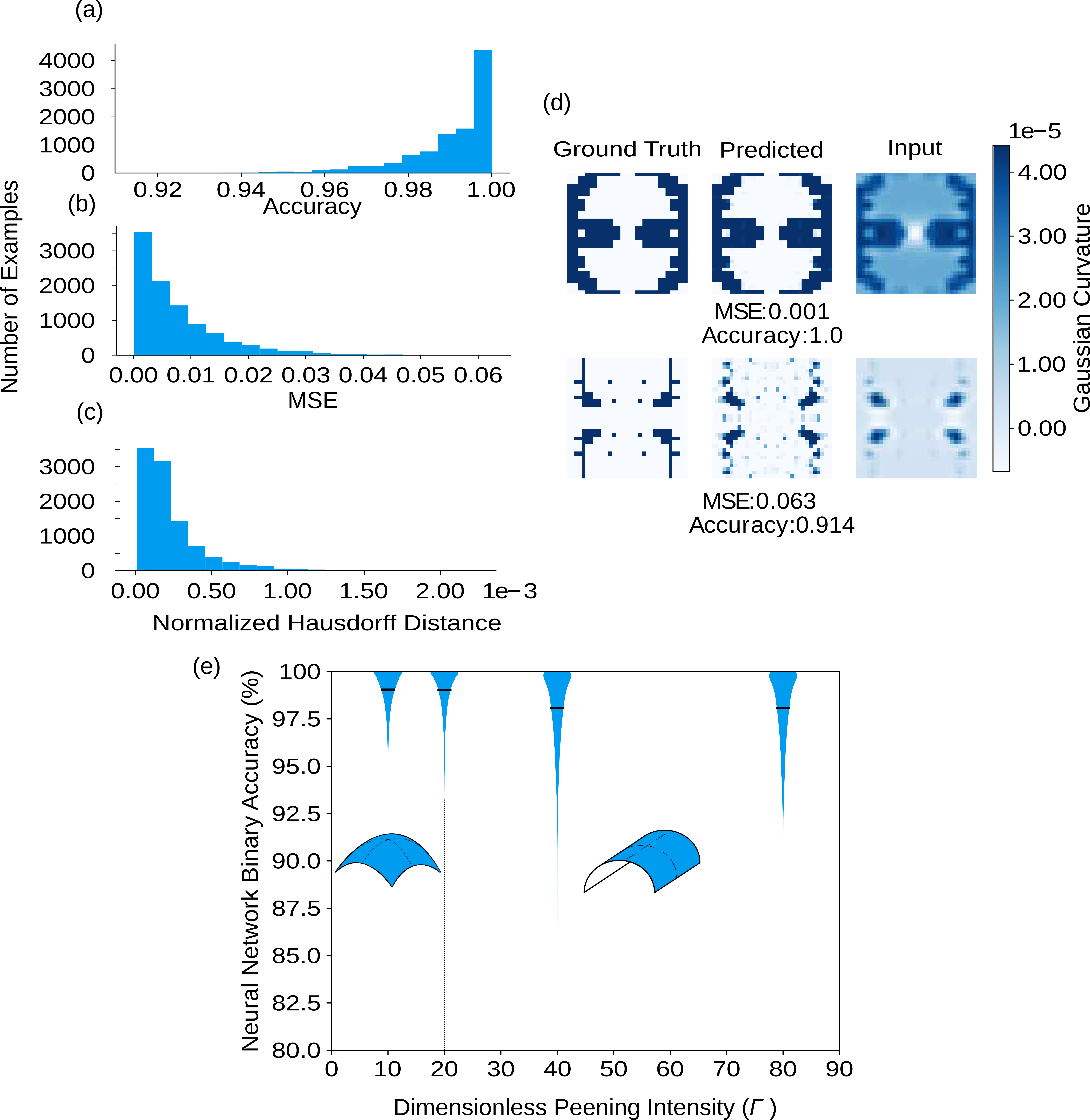}
     \caption{\label{fig:results}Distribution of (a) binary accuracy and (b) mean squared error over the test set; (c) Distribution of Hausdorff distance in the test set, normalized by the mesh bounding box diagonal. The fact that (a),(b) and (c) show high accuracy, low MSE and low Hausdorff distance for all test set examples, validates the generalization of the model beyond the training set; (d) The normalized confusion matrix demonstrates the presence of small percentages of false negatives and false positives; (e) A random selection of the best and worst examples from the test set, presenting the ground truth reference (left), the predicted pattern (middle), and the input Gaussian curvature image (right);  (f) With the increase of intensity, the plates are proven to deform in a cylindrical mode, rather than spherical \cite{FAUCHEUX2018135}. The violin plot presents the distribution of the test set binary accuracy for various dimensionless peening intensities, with highlighted mean values. We observe a slight increase in the accuracy standard deviation in the cylindrical domain.}
         
     \centering
\end{figure}

The presented results were computed on a test set of 10,000 examples simulated with $\Gamma=20$. Fig. \ref{fig:results}a shows that the binary accuracy evaluated on the test set averages 98.8\%. This accuracy metric is widely used for image classification, and in this context it represents the percentage of elements correctly classified. Fig. \ref{fig:results}b shows that the majority of the test examples exhibit a MSE close to zero.

The Hausdorff distance \cite{hausdorff} computed between the mesh nodes of the deformed plate under the peening pattern produced by the NN and the reference peening pattern measures the deviation of the resulting shape from the target shape. Both meshes are defined in the same simulation reference frame, with the centre node at the origin. Normalizing this distance by the diagonal of the mesh bounding box offers a dimensionless metric proportional to the size of the deformed plate. Fig. \ref{fig:results}c shows the distribution of the normalized Hausdorff distance for the test set. It further demonstrates the performance of the model, with an average of 0.02\%. Fig. \ref{fig:results}d presents the normalized confusion matrix computed on the test set. It shows that the percentages of false positives and false negatives are small. These results are complemented with a satisfactory F1-score of 0.981.

Fig. \ref{fig:results}e shows a sample of the best and worst cases of patterns predicted by the NN as compared to ground truth for a given input Gaussian curvature.

The relationship between the applied peening intensity and the simulated curvature of the plate is non-linear. For a uniformly peened square plate, at low intensity ($\mathit{\Gamma}<20$) it adopts a spherical deformation, whereas at high intensity ($\mathit{\Gamma}>20$) it adopts a cylindrical deformation \cite{FAUCHEUX2018135}. We thus check how the ability of the NN to predict the right pattern is affected by the peening intensity. To do so, we train the NN four times with four datasets obtained at four peening intensities. During this process, the hyper-parameters and network architecture remain unchanged.
Fig. \ref{fig:results}f represents the distribution of binary accuracy for various peening intensities (i.e., different values of dimensionless bending moment $\mathit{\Gamma}$). Overall, the accuracy decreases slightly with the increase in the peening intensity, but remains high even for very large intensities. All instances of the neural network converged to a low loss value, without exhibiting signs of overfitting.

\pagebreak
\section{Conclusion and outlook}
\label{sec5}
The main contribution of this work is the demonstration that the inverse problem of peen forming can be represented as a visual pattern recognition problem, by encoding the geometry of the target shape in terms of Gaussian curvature. This approach solves the problem via a neural network, a proven technique for pattern recognition tasks.

The limitations of this model lie in the need for a large dataset that represents the various possible target shapes that can be achieved using a given treatment intensity. Proving the generalization of such model on any possible shape is an open problem.

This work constitutes the first step towards the long-term goal of developing closed-loop peen forming by incorporating neural networks and the feedback provided by the geometric inspection data from a 3D scanner to compensate for deviations caused by unknown residual stresses in peen formed parts.
 The data-driven nature and differentiability of NNs are the central features that demonstrate their potential to adapt to new examples, therefore, correct their prediction errors based on new data.

Since our developed method makes use of the bilayer framework, it would easily be adaptable to a broad range of problems which can also be simulated with bilayers such as biological growth \cite{van_rees_growth_2017,haas_pectin_2020},  active materials \cite{thermorph} or solvent absorption \cite{pezzulla_geometry_2016}.

\section*{Acknowledgements} 
This work was funded by Aerosphere Inc., the Fonds de recherche du Québec – Nature et technologies and the ministère de l'Économie, de la Science et de l'Innovation [Funding reference number LU-210888].\\

\textcopyright 2020. This manuscript version is made available under the CC-BY-NC-ND 4.0 license http://creativecommons.org/licenses/by-nc-nd/4.0/ 



\bibliographystyle{elsarticle-num}

\bibliography{sample}

\end{document}